\documentstyle[aps,prl,preprint,floats,epsfig]{revtex}

\providecommand{\btopsikzero}{\mbox{$ { B}^0\rightarrow 
{ J}/\psi  K^0$}}

\providecommand{\btodstplusdstminus}
{\mbox { ${ {  B^0} \rightarrow  D^{*+}  D^{*-} } $}}
\providecommand{\dstplusdstminus}
{\mbox { $  D^{*+}  D^{*-}  $}}

\providecommand{\BZ}{\mbox{${ B}^0$}}

\providecommand{\BZBZBAR}{\mbox{ $ { B^0} \overline { B}^0 $ }}

\providecommand{\DBAR}{\mbox{$\overline{ D}$}}
\providecommand{\DP}{\mbox{${ D}^+$}}

\providecommand{\DZ}{\mbox{${ D}^0$}}

\providecommand{\CLEO}{\mbox{CLEO}}
\providecommand{\ALEPH}{\mbox{ALEPH}}

\providecommand{\UFOURS}{\mbox{$\Upsilon(4{ S})$}}

\providecommand{\epem}{\mbox{${ e}^+{ e}^-$}}

\providecommand{\invfb}    {\mbox{$ {\mathrm fb^{-1}} $}}

\providecommand{\cleoii}{\mbox{CLEO II}}
\providecommand{\cleoiiv}{\mbox{CLEO II.V}}

\providecommand{\dptokpipi}{\mbox{${ D^{+}}\rightarrow \KPIPI $}}

\providecommand{\dele}{\mbox{$\Delta { E} $}}

\providecommand{\mb}{\mbox{$M_{ B}$}}

\providecommand{\lbysigl}{\mbox{${\mathrm L}/\sigma({\mathrm L})$}}
\providecommand{\sigl}{\mbox{${\sigma({\mathrm L})}$}}
\providecommand{\chisqm}{\mbox{$\chi^2_{ M}$}}

\providecommand{\slowpi}{\mbox{$\pi_{\mathrm s}$}}

\providecommand{\dstplusdstminus}{\mbox { ${  D^{*+}  D^{*-} } $}}
\providecommand{\PHI}{\mbox{$\phi$}}
 
\def\measbr{({6.2}^{+4.0}_{-2.9} \pm 1.0) \times 10^{-4}}
\def\ssmeasbr{[{6.2}^{+4.0}_{-2.9}  ({\mathrm stat}) \pm 1.0  ({\mathrm syst}) ]\times 10^{-4}}
\def\estbkgd{0.31\pm 0.10}
\def\flucprob{1.1 \times 10^{-4}} 
\def\commsys{31\%}

\def\Dstarplandmi{{ D}^{*+,-}}  

\def\MeV{{\mathrm MeV}}
\def\piplpimi{\pi^+\pi^-}
\def\Dpl{{ D}^+}
\def\Dmi{{ D}^-}
\def\pizero{{\pi}^0}
\def\Kpl{{ K}^+}
\def\Kmi{{ K}^-}
\def\Dzero{{ D}^0}

\def\Dstarplus{{ D}^{*+}}
\def\Dstarminus{{ D}^{*-}}
\def\DsDs{{ D}^{*+}{ D}^{*-}}
\def\B0DsDs{{ B}^0\rightarrow\DsDs}

\def\KS{{ K}^0_{\mathrm S}}
\def\Br{{\cal B}}
\def\bbbar{{ B}\overline{ B}}
\def\ccbar{{c}\overline{c}}
\def\Dstar{{ D}^{*+}}
\def\slowpi{\pi^+_{\rm s}}
\def\pizeroslow{\pi^0_{\rm s}}
\def\dstoDzeropi{\Dstar\rightarrow{ D}^0\slowpi}

\def\zp{({ D}^0\pi^+_{\rm s})({ D}^-\pi^0_{\rm s})}
\def\zz{({ D}^0\pi^+_{\rm s})(\overline{ D}^0\pi^-_{\rm s})}

\def\d0pi{({ D}^0\pi^+_{\rm s})}
\def\dpi0{}

\def\Dzerotokpi{{ K}^-\pi^+}
\def\Dzero_kpipizero{{ K}^-\pi^+\pi^0}
\def\DzeroTo_K3pi{{ K}^-\pi^+\pi^+\pi^-}
\def\Dzeroto_Kspipi{{ K}^0_{\rm S}\pi^+\pi^-}
\def\DzerotoK0s2PIpizero{{ K}^0_{\rm S}\pi^+\pi^-{\pi^0}}

\def\dptokpipi{{ K}^-\pi^+\pi^+}
\def\dptokspi{{ K}^0_{\rm S}\pi^+}
\def\dplustokspipizero{{ K}^0_{\rm S}\pi^+\pi^0}
\def\dptokspipipi{{ K}^0_{\rm S}\pi^+\pi^+\pi^-}
\def\dptophipi{\phi\pi^+}
\def\dptophipipizero{\phi\pi^+\pi^0}

\begin{document}

\preprint{\tighten\vbox{\hbox{\hfil CLNS 98/1589}
                        \hbox{\hfil CLEO 98-17}
}}

\title{First Observation of the Decay {\mathversion{bold}$\btodstplusdstminus$}}  

\author{CLEO Collaboration}
\date{\today}

\maketitle
\tighten

\begin{abstract} 

We have observed four fully reconstructed $\B0DsDs$ candidates 
in 
5.8 million   $\Upsilon(4{S}) \rightarrow \bbbar$ decays recorded
with the \CLEO\ detector.
The background 
is estimated to be 
$\estbkgd$ events.
The probability that the background could produce four or more
signal candidates with the observed distribution among $\Dstarplus$ and $\Dstarminus$
decay modes is $\flucprob$.
The measured decay rate,
${\mathcal B}(\B0DsDs) = \ssmeasbr $,
is large enough for this decay mode to be of interest
for the measurement of a time-dependent CP asymmetry.

\end{abstract}
\newpage

{
\renewcommand{\thefootnote}{\fnsymbol{footnote}}

\begin{center}
M.~Artuso,$^{1}$ E.~Dambasuren,$^{1}$ S.~Kopp,$^{1}$
G.~C.~Moneti,$^{1}$ R.~Mountain,$^{1}$ S.~Schuh,$^{1}$
T.~Skwarnicki,$^{1}$ S.~Stone,$^{1}$ A.~Titov,$^{1}$
G.~Viehhauser,$^{1}$
% ...
\end{center}
 
\small
\begin{center}
$^{1}${Syracuse University, Syracuse, New York 13244}\\
% ...
\end{center}

\setcounter{footnote}{0}
}
\newpage

\par The Cabibbo-suppressed decay $\B0DsDs$
is a promising channel for searches 
of CP violation in \BZ\ meson decays at  future 
$B$ 
factories~\cite{babartdr,belletdr}. 
Within the framework of the Standard Model, 
the proper time-dependent CP asymmetry in the decay \btodstplusdstminus\ 
could provide a measurement of the angle $\beta$ of the unitarity triangle~\cite{pdg}
in the same way as the well-known decay {\btopsikzero}~\cite{babartdr,belletdr}.
The final state \dstplusdstminus\ may be an admixture of CP-even and 
CP-odd states which could complicate such a measurement. However, 
the two CP components 
can be disentangled using  angular correlations in the final 
state~\cite{dunietz}, and 
estimates based on the heavy quark limit indicate that the 
dilution of the asymmetry from the two different CP states  
is small~\cite{aleksan}.
The decay amplitude for the process $\B0DsDs$ 
is expected to be dominated by 
the decay $ \overline b\rightarrow \overline c W^+;~W^+ \rightarrow c\overline d$.
The branching fraction 
for this process  can be estimated from the measured rate~\cite{dsdrefs} of the 
Cabibbo-favored decay 
${ B}^0\rightarrow{ D}_{ s}^{*+}{ D}^{*-}$ and is
$  \Br(\B0DsDs ) \approx  
    ({f_{ D^*}} / {f_{{ D}^{*}_{ s}}}
    )^2 \,
    \tan^2\theta_{\rm C}\,
  \Br({ B}^0\rightarrow{ D}_{ s}^{*+}{ D}^{*-})
 \approx 0.1\% $, where the $f_X$ are the decay constants
and $\theta_{\rm C}$ is the Cabibbo angle.

\par The \CLEO~\cite{cleo}
and \ALEPH~\cite{aleph} collaborations have 
searched for the $\B0DsDs$  decay   and have 
reported 90\% CL  upper limits 
on the branching fraction
of $22 \times 10^{-4}$ and $61 \times 10^{-4}$, respectively.
In this Letter 
we report on  the first observation of the decay $\B0DsDs$ and a measurement of its 
decay rate. This measurement supersedes the previous \CLEO\ search~\cite{cleo}.

\par The data were recorded at the Cornell Electron Storage Ring (CESR)
with two configurations of the
\CLEO\ detector, called {\cleoii}~\cite{kubota} and {\cleoiiv}.
In  the  \cleoiiv\ configuration, the innermost wire chamber was replaced with a
precision three-layer silicon vertex detector (SVX)~\cite{svx}.
Each layer of the SVX  is equipped with readout on both sides 
providing precise measurements of 
the $\phi$ and $z$ coordinates of the charged particle trajectory. 
(The $z$-axis of the \CLEO\ cylindrical coordinate system 
is coincident with the $e^+$ beam direction.)
The results in this Letter are based upon an integrated luminosity of 
$3.14\,(2.46)\, \invfb$ of \epem\ 
data recorded at the $\Upsilon(4S)$ energy and 
$1.57\,(1.26)\, \invfb$ recorded $60\,\MeV$ below the $\Upsilon(4S)$ energy 
with the \cleoii\ ({\cleoiiv}) configuration.
The Monte Carlo simulation of the  \CLEO\ detector response was based upon 
GEANT~\cite{GEANT}. Simulated events
for the \cleoii\ and \cleoiiv\ configurations
were processed in the same manner as the data. 

\par  Candidates for the decay $\B0DsDs$ with the 
subsequent decays $\Dstarplus \rightarrow { D}^0 \slowpi$ and 
$\Dstarplus \rightarrow \Dpl \pizeroslow$ were selected.
The ${ D}^0$ and $\Dpl$ mesons were reconstructed in the eleven decay modes
listed in Table~\ref{tab:dbratio}. 
In this Letter, ``$D$'' refers to both \DZ\ and \DP\ mesons, 
and ``$\pi_{\rm s}$'' refers to the slow pion produced in $\Dstar$ decay. 
In addition, reference to charge conjugate states
is implicit unless explicitly stated.
The charged track candidates from $\Dstar$ and $D$ meson decays
were required to originate near the $\epem$ 
interaction point.
Charged kaons and pions were distinguished using the 
charged particle's measured specific
ionization ($dE/dx$) and time of flight across the tracking volume.
We required that the 
$dE/dx$ and time-of-flight 
information was consistent
with the $D$ daughter hypotheses of the particular $D$ meson decay mode.
Charged tracks and $\KS$ candidates forming a $D$ candidate were required to 
originate from a common vertex.
The $\KS$ candidates were selected through their decay into $\piplpimi$ mesons.
The decay point of the $\KS$ candidate was required to be 
displaced from the $\epem$ interaction point
and at least one 
daughter pion was required to be inconsistent with originating at 
the interaction point.
Neutral pions were reconstructed from photon pairs detected in the 
electromagnetic calorimeter.
The photons were required to have an energy of at 
least 30(50) MeV  in the barrel(endcap) region, and
their invariant  mass was required to be within three
standard deviations of the nominal $\pizero$ meson mass~\cite{pdg}.
The $\pizero$ momentum was required to 
be at least 70(100) MeV for $\Dstar$($D$) daughters.
To reduce backgrounds, we accepted 
both $\zz$ and $\zp$ combinations 
but not $(\Dpl\pizeroslow)(\Dmi\pizeroslow)$.
A fit constraining the mass of each $\Dstar$ candidate to the nominal
value~\cite{pdg} improved the $\Dstar$ momentum resolution by 14\% in simulated events.

\begin{table} 
  \caption{\label{tab:dbratio}The \DZ\ and \DP\  meson decay modes used in this
           analysis and their 
           branching fractions~\protect\cite{pdg}. The branching fractions 
            ${\cal B}(\KS \rightarrow \piplpimi)$ 
           and ${\cal B}(\PHI \rightarrow \Kpl\Kmi)$ are included for
           the modes containing $\KS$ or $\PHI$ mesons.
            }
  \begin{tabular}[htbp]{|l|c||l|c|}
    \multicolumn {2}{|c||}{\DZ\ Decay Modes}    &
    \multicolumn {2}{c|}{\DP\ Decay Modes}        \\ \hline
                & Branching     &               & Branching      \\
    Decay Mode  & Fraction (\%) &  Decay Mode   & Fraction (\%)     \\ \hline 
$\Dzerotokpi$            & $3.85 \pm 0.09$   & $\dptokpipi$        & $9.0 \pm 0.6$ \\ \hline   
${ K}^-\pi^+\pi^0$       & $13.9 \pm 0.9$    & $\dptokspi$         & $1.0 \pm 0.1 $\\\hline
$\DzeroTo_K3pi$          & $7.6 \pm 0.4$     & $\dplustokspipizero$& $3.3 \pm 1.0$ \\\hline  
$\Dzeroto_Kspipi$        & $1.9 \pm 0.1$     & $\dptokspipipi$     & $2.4 \pm 0.3$\\\hline  
$\DzerotoK0s2PIpizero$   & $3.4 \pm 0.4$     & $\dptophipi$        & $0.30 \pm 0.03$ \\\hline
                         &                & $\dptophipipizero$  & $1.1 \pm 0.5$  \\\hline
Total                    & $30.6 \pm 1.3$ & Total            & $17.1\pm 1.6$   \\ 
  \end{tabular} 
\end{table}
\par The $\B0DsDs$ candidates were selected by means of four
observables.
The first observable, {\chisqm},  measured
the deviation of each $D$ and $\Dstar$ candidate 
from the nominal mass ($M^{\rm n}_i$) and mass difference ($\Delta {M}^{\rm n}_i $),
respectively,
 \begin{equation}
  \label{eqn:chi2m}
 \chi^2_M \equiv \sum_{i=1,2} 
\left[
\left(\frac {M_i - M^{\rm n}_i}{\sigma({M_i})}\right)^2 + 
\left(\frac {\Delta {M_i} - \Delta {M}^{\rm n}_i}
            {\sigma(\Delta {M_i})}\right)^2
\right]
,
\end{equation}
\noindent where $\sigma(M_i)$ and $\sigma(\Delta {M_i})$ are the average resolutions 
in the reconstructed $D$ candidate mass $M_i$ and the mass difference
$\Delta {M_i} \equiv M_i(\Dstar) - M_i$, respectively, and 
$i=1,2$ corresponds to the $\Dstarplandmi$ and the $D$,\DBAR\ daughters.
If an event had more than one $\B0DsDs$ candidate, then the candidate
with the lowest \chisqm\ was chosen.
The second observable, {\lbysigl}, is 
the significance of the projected three-dimensional 
distance L between the reconstructed 
$D$ and \DBAR\ meson decay vertices,
$$
{\mathrm L} = (\vec {V}_{ D} - \vec {V}_{\overline{ D}}) \cdot
 \frac {(\vec {p}_{ D} - \vec {p}_{\overline{ D}})}
       {|\vec {p}_{ D} - \vec {p}_{\overline{ D}}|}
\ \ ,
$$
\noindent where $\vec{p}_D$ and 
$\vec{V}_D$ are the momentum and decay vertex position of
the $D$ candidate, respectively, and
\sigl\ was calculated from the 
covariance matrices of the $D$ and \DBAR\
tracks resulting from the vertex fits of the $D$ daughters.
This observable exploits the relatively long $\Dpl$ meson lifetime
and the precise decay vertex resolution available in {\cleoiiv}.
The difference between the energy of the $B^0$ candidate
and the beam energy, $\Delta E \equiv E(\Dstarplus) + E(\Dstarminus) - E_{\rm beam}$,
is the third observable.
In simulated \btodstplusdstminus\ decays,
the $\Delta E$ resolution is $8.0\, \MeV$.
The fourth observable is the beam-constrained \BZ\ candidate mass 
$M_{ B} \equiv \sqrt{E_{\rm beam}^2-{\vec{p}_{ B}}^{\  2}}$ where
$\vec{p}_{ B}$ is the momentum of the $B$ candidate.
The resolution of $M_{ B}$, dominated by the beam energy spread, was measured to 
be $2.5\, {\mathrm MeV}$~\cite{BtoDK}.

\par The selection criteria for these four observables
were optimized for signal significance
using simulated signal and background events.
We also checked the optimization using background distributions estimated from
the data by combining $D$ candidates with  $\slowpi$ and $\pizeroslow$ candidates
with the momentum vector direction reversed and found similar results.
The optimal criteria determined were 
$\chi^2_{ M} < 10$, 
${\mathrm L}/\sigma({\mathrm L}) > 0 $ for
the $\zp$ candidates in the \cleoiiv\ data only, 
$|\Delta E| < 20 \ \MeV$ and 
$|\Delta M_{ B}| \equiv |M_{ B} - M_{B}^{\rm nom}| < 6.25 \
{\mathrm MeV}$ where $M_{B}^{\rm nom}$ is the nominal \BZ\ meson mass~\cite{pdg}.

\par With these criteria, the reconstruction efficiency for
each $\Dstar$ and $D$ decay channel was measured from simulated $\B0DsDs$
decays. 
Important  issues 
 in $\B0DsDs$ reconstruction are the ability to reconstruct
the trajectory of charged slow pions $\slowpi$ that populate the momentum range 
from 60 to 190 MeV and the accurate determination of their reconstruction
efficiency.
The track-finding algorithm used for these results was optimized
for the \cleoii\ but not the \cleoiiv\ configuration. As a result,
the ratio of the $\dstoDzeropi$ reconstruction efficiency in the
\cleoiiv\ data to that in the \cleoii\ data is ($65\pm 6$)\%
due to the reduced $\slowpi$ reconstruction efficiency.
We corrected the $\Dstar$ reconstruction efficiency for differences
in the inclusive $\Dstar$ meson yields
between data and simulation in this momentum range 
using the measured inclusive $\Dstar$ production spectrum
in $\Upsilon(4{S}) \rightarrow \bbbar$ events~\cite{moneti}.
Including
the $\Dstar$ and $D$ daughter branching fractions,
the overall reconstruction efficiency  was 
${\cal E} = (10.08 \pm 1.10) \times 10^{-4}$. An appropriate
figure of merit is the single event sensitivity, defined as 
{$[2N(\bbbar)f_{00}{\cal E}]^{-1}$,}
where $N(\bbbar)$ is the number of $\bbbar$ pairs and $f_{00} = 0.48 \pm 0.04$
is the fraction of \UFOURS\ decays to {\BZBZBAR}~\cite{f00}.
Our sample of 
$3.3 \times 10^6$ ($2.5 \times 10^6$) $\bbbar$ pairs in the 
\cleoii\ ({\cleoiiv}) data gives a single
event sensitivity for $\B0DsDs$ of 
$(1.8\pm 0.3)\times 10^{-4}$.

\par
\begin{figure}[htbp]
  \begin{center}
    \epsfysize=15.0cm 
     \epsfbox{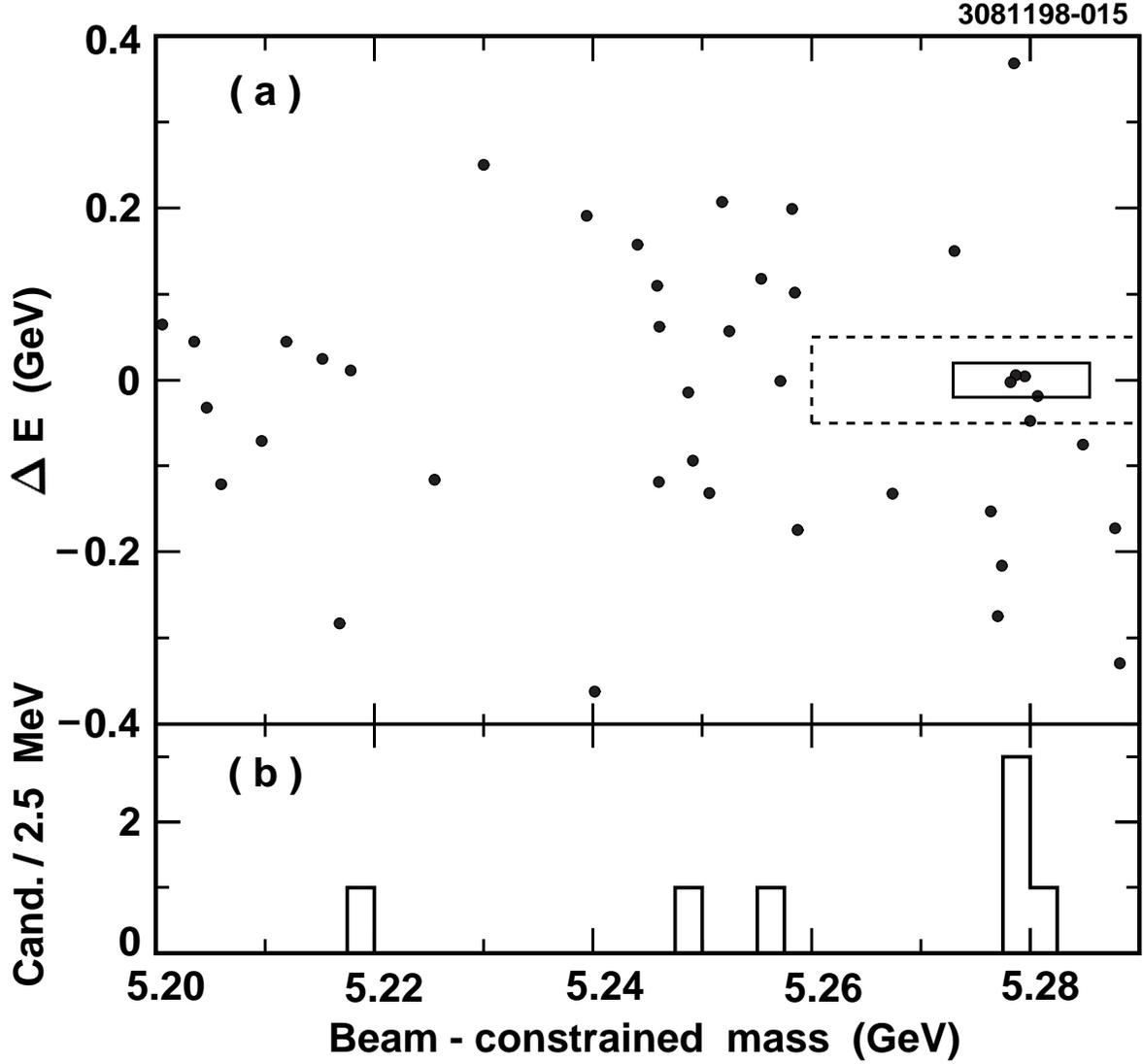}
 \caption{\label{fig:4s-dele-mb-allmodes}
  (a) The \dele\ {\it vs} the beam-constrained mass distribution for  
  all $\B0DsDs$ candidates in the data taken on the \UFOURS\ resonance.
  The signal region is indicated by the box with the solid line.
  The area outside the dashed line is the grand sideband (GSB).
  There are four candidates in the signal region and a total
  of 41 candidates in the entire distribution.
  (b) The beam-constrained mass distribution for $\B0DsDs$ candidates
      satisfying  $|\Delta E| < 20 \ {\rm MeV}$.
 }    
  \end{center}
\end{figure}

 We used two independent methods to estimate
the contributions of the background  to the signal region,
defined as 
{$|\Delta E| < 20 \ \MeV$} and 
{$|\Delta M_{ B}| < 6.25 \ {\mathrm MeV}$.}
In Method 1, we scaled the number of candidates in a grand
sideband (GSB) to estimate the background contribution to the signal
region.
The GSB is defined
by the regions
{($50 < |\Delta E| < 400 \ \MeV$} and  
{$5.20 < M_{ B} < 5.29 \ {\mathrm GeV}$)}
or 
{($ |\Delta E| < 400 \ \MeV$} and 
{$5.20 < M_{ B} < 5.26 \ {\mathrm GeV}$)}
and is indicated in Figure~\ref{fig:4s-dele-mb-allmodes}(a) 
as the area outside the dashed line.
The scale factor for the GSB events is 
the ratio of the area of the signal region to area of the GSB.
The estimated background 
contribution to the signal region is $0.261 \pm 0.043$ events
from Method 1.
In principle, the background contribution determined from the GSB
slightly overestimates the actual background due to 
{${ B} \rightarrow \Dstarplus\Dstarminus X_{ s,d}$} decays that
are kinematically forbidden to 
populate the signal region but may be present in the \dele\ or
\mb\ sideband regions. 
This overestimation is negligible as discussed below.

\par For Method 2, we decomposed the background into four classes
and estimated the contribution of each class separately.
The dominant background class is composed of 
random combinations of $\Dstarplus$ and $\Dstarminus$
candidates in which either one or both candidates is ``fake''; that is, they are
not composed of the daughters of an actual $\Dstar$ decay. 
The other background classes comprise combinations in which the $\Dstarplus$ and
$\Dstarminus$ candidates arise from actual $\Dstarplus$ and $\Dstarminus$ meson decays
that are roughly back-to-back. The contributing processes
are 
(1) {${ e}^+{ e}^- \rightarrow \ccbar$} with
   {$c \rightarrow \Dstarplus$} and 
   {$\overline{c} \rightarrow \Dstarminus$,}
(2) {$\Upsilon(4{S}) \rightarrow \bbbar$} with
   {${ B} \rightarrow \Dstarplus X$} and
   {$\overline{ B} \rightarrow \Dstarminus Y$},
and 
(3) {${ B} \rightarrow \Dstarplus\Dstarminus X_{ s,d}$} where
$X_{ s,d}$ represents either a strange or non-strange meson from the
decay of an orbitally- or a radially-excited $D$ meson or non-resonant
$\Dstar X$ production. 
\par We estimated the combinatorial background from data with
explicit fake $\Dstarplus$ candidates formed by replacing
$M_1^{\mathrm n}$ in \chisqm\ in equation (\ref{eqn:chi2m}) with 
$M_1^{\mathrm n} + 6\sigma(M_1)$ or 
$M_1^{\mathrm n} - 6\sigma(M_1)$. 
We first formed a sample of fake $\Dstarplus$ candidates
combined with standard $\Dstarminus$ candidates.
Similarly, we formed a sample of fake $\Dstarplus$ candidates
combined with fake $\Dstarminus$ candidates. 
The combinatorial background 
can be derived from these samples and contributes an estimated 
$0.304 \pm 0.040$ events when scaled to the signal region.

\par The contribution of the process (1)
 {${ e}^+{ e}^- \rightarrow \ccbar$},
   {$c \rightarrow \Dstarplus$},
   {$\overline{c} \rightarrow \Dstarminus$}
was estimated from the data taken
60 MeV below the \UFOURS\ after subtracting the combinatorial background
using the method  described above. The estimated rate of 
 {${ e}^+{ e}^- \rightarrow \ccbar$},
   {$c \rightarrow \Dstarplus$},
   {$\overline{c} \rightarrow \Dstarminus$}
was corrected for the relative cross section and luminosity and
scaled to the area of the signal region. The estimated number
of events in the signal region from this process was $0.039 \pm 0.030$.

\par The contributions of processes (2) and (3) 
from {$\Upsilon(4{S}) \rightarrow \bbbar$}
were estimated from a sample of simulated events approximately ten times 
the data sample.
The process (2) 
 {$\Upsilon(4{S}) \rightarrow \bbbar$},
   {${ B} \rightarrow \Dstarplus X$},
   {$\overline{ B} \rightarrow \Dstarminus X$}
was estimated to contribute 
$0.024 \pm 0.003$ events to the signal region.
The contribution of 
 {${ B} \rightarrow \Dstarplus\Dstarminus X_{ s}$} was
determined to be negligible
assuming $\Br(B \rightarrow \Dstarplus \Dstarminus X_s) = 1.8\%$~\cite{roy}. 
The contribution of
 {${ B} \rightarrow \Dstarplus\Dstarminus X_{ d}$} 
was estimated from a simulation of 
 {${ B} \rightarrow \Dstarplus { D}^{**-}_{\mathrm sim}$} 
in which the ${ D}^{**-}_{\mathrm sim}$ pseudo-particle
had a mass of 2420 ${\mathrm MeV}$ and a width 
of 400 ${\mathrm MeV}$ and was forced to decay
to the $\Dstarminus \pi^0$ final state. 
The reconstructed $\Dstarplus\Dstarminus$ combination from this process tends to
have  $\Delta E < -M_{X_d}$ so the probability of reconstructing it
with {$ |\Delta E| < 400 \ \MeV$} and 
{$5.20 < M_{ B} < 5.29 \ {\mathrm GeV}$}
is twelve times smaller than that for the signal process.
Assuming 
$\Br({ B} \rightarrow \Dstarplus\Dstarminus X_{ d}) \approx \Br(\B0DsDs)$, this 
contribution to the signal region is negligible.
The estimated contribution to the signal region from the sum of
all backgrounds is  $0.367 \pm 0.051$ events
for Method 2.

\par The background rates obtained from these two statistically independent
methods were averaged to yield the estimated background contribution
to the signal region of  
$0.306 \pm 0.033  ({\mathrm stat}) \pm 0.094 ({\mathrm syst})$.
The $\commsys$ systematic uncertainty arises from the uncertainty in the
shapes of the \dele\ and \mb\ distributions of the background.
The systematic uncertainty was taken to be the difference in
the scale factor when these distributions were fitted
with second- and first-order polynomials, respectively, instead of a zeroth-order polynomial.

\par 
The distribution of the 
41 candidates passing the selection criteria in the \UFOURS\
data sample in the \dele\ {\it vs} \mb\ plane  is
shown in Figure~\ref{fig:4s-dele-mb-allmodes}. 
There are four candidates in the signal region.
The observed number of candidates and the estimated background
for the $\zz$ and $\zp$ submodes are listed in Table~\ref{tab:bkgd-estimate}.
Also listed in Table~\ref{tab:bkgd-estimate} is the probability that
a fluctuation of the estimated background could produce the observed 
number of signal candidates
or more in each submode. 
The calculation of the background fluctuation probability
assumes that the statistical uncertainty in the background 
in the two submodes is uncorrelated
and that the systematic uncertainty in the background 
is fully correlated between the submodes.
Integrating over all background levels,  assuming that
the number of background events is normally distributed about its
central value for each submode~\cite{cousins},
we find that the combined probability that the
estimated background could produce the observed number of signal candidates
or more in the two submodes is $\flucprob$.

\par  The branching fraction of $\B0DsDs$ was calculated using a maximum likelihood
technique that took into account 
the signal efficiency and estimated background contribution
for each $D$ decay mode in the \cleoii\ and \cleoiiv\ data samples.
The branching
fraction was determined to be 
$\Br(\B0DsDs) = \measbr$, where the first error is statistical
and the second is systematic. The systematic uncertainty is dominated
by the uncertainties  in ${\cal E}$ ($13.3\%$) and $f_{00}$ ($8.3\%$).
The product branching fraction,
using the $\Dstar$ and $D$ decay modes as in this Letter,
$\Br(\B0DsDs)\times 
 \sum \Br(\Dstar \rightarrow D\pi)\times 
      \Br(D\rightarrow X) \approx 4\times 10^{-5}$,
is comparable to that of the {\btopsikzero} decay mode,
$\Br(B^0 \rightarrow J/\psi K^0) \times 
 \sum \Br(J/\psi \rightarrow \ell^+\ell^-)
 \times \Br(K^0\rightarrow\KS\rightarrow\pi^+\pi^-) \approx 3.6\times 10^{-5}$.
With careful optimization of the charged track reconstruction efficiency, in
particular that of charged slow pions, exclusively reconstructed
$\B0DsDs$ decays could permit a complementary measurement of
the angle $\beta$ of the unitarity triangle at future $B$ factories.
  \begin{table}
    \caption{\label{tab:bkgd-estimate}
             The efficiency, observed number of 
             candidates  and estimated number of  background events
             in the $\zz$  and $\zp$ decay submodes.
             The reconstruction efficiency ${\cal E}$ includes
             the $D$ and $\Dstar$ daughter branching fractions. 
             The row labeled ``All ({\dele},{\mb})'' 
             is the total number of $\B0DsDs$ candidates
             in each submode in the  $5.20 < M({\rm B}) < 5.29 \ {\rm GeV}$ and
             $|\Delta E| < 400 \ {\rm MeV}$ region.
             The row labeled ``Signal region'' contains the 
             observed number of signal candidates  in the 
             $|\Delta E| < 20 \ \MeV$ and 
             $|\Delta M_{ B}| < 6.25 \ {\mathrm MeV}$ region.
             ``Bkg. Meth. 1'' and ``Bkg. Meth. 2'' are the number of background
             events in the signal region 
             estimated using the two independent methods described in the text.
             The sixth row contains the 
             average estimated number of background events in the signal
             region. Only statistical uncertainties are 
             included for the upper six rows.
	     The calculation of the background fluctuation probability
             ${\bf P}$ is described in the text.
             }
     \begin{tabular}[ht]{|l|c|c||c|}
                             & $({ D}^0\pi^+_{\rm s})$     & $({ D}^0\pi^+_{\rm s})$          &  \\
                          &$(\overline{ D}^0\pi^-_{\rm s})$ &$({ D}^-\pi^0_{\rm s})$          & Total \\
\hline 
\hline 
${\cal E}\times 10^4$        &$6.06\pm 1.02$       &$4.02\pm 0.40$& $10.08\pm 1.10$\\
\hline 
All({\dele},{\mb})           &  13                 &  28            & 41 \\
Signal Region                &   2                 &   2            &  4 \\
\hline 
\hline 
Bkg. Meth. 1                 &$0.080\pm 0.024$     &$0.181\pm 0.036$& $0.261\pm 0.043$ \\
Bkg. Meth. 2                 &$0.091\pm 0.024$     &$0.275\pm 0.044$& $0.367\pm 0.051$ \\
\hline 
Average Bkg.                 &$0.085\pm 0.017$     &$0.219\pm 0.028$& $0.306\pm 0.033$ \\
\hline 
${\bf P}$ &$3.85\times 10^{-3}$ & $2.24\times 10^{-2}$ & $1.10\times 10^{-4}$ \\
     \end{tabular}
   \end{table}
\par In conclusion,
we have fully reconstructed four $\B0DsDs$ candidates with a total estimated background
of $\estbkgd$ events
in $5.8 \times 10^6$ {$\Upsilon(4{S}) \rightarrow \bbbar$} decays. 
The probability that the estimated background
could fluctuate to the observed number of signal candidate events or
more is $\flucprob$.
The branching fraction is measured to be
$\Br(\B0DsDs) = \ssmeasbr$ and 
is in agreement with the expected rate. This rate
suggests that
this decay  could provide an avenue for the measurement of
the angle $\beta$ of the unitarity triangle.

\par We gratefully acknowledge the effort of the CESR staff in providing us with
excellent luminosity and running conditions.
We also thank the staffs of our institutions in providing us
with a superbly performing detector.
This work was supported by 
the National Science Foundation,
the U.S. Department of Energy,
the Research Corporation,
the Natural Sciences and Engineering Research Council of Canada, 
the A.P. Sloan Foundation, 
the Swiss National Science Foundation, 
and the Alexander von Humboldt Stiftung.  

\end{document}